\documentstyle[emulateapj,epsfig]{article}

\begin{document}

\title{Multi-Color Observations of the GRB~000926 Afterglow}
\author{P.~A.~Price\altaffilmark{1,2},
F.~A.~Harrison\altaffilmark{1},  
T.~J.~Galama\altaffilmark{1}, 
D.~E.~Reichart\altaffilmark{1},
T.~S.~Axelrod\altaffilmark{2},
E.~J.~Berger\altaffilmark{1},
J.~S.~Bloom\altaffilmark{1},
J.~Busche\altaffilmark{3},
T.~Cline\altaffilmark{4},
A.~Diercks\altaffilmark{1},
S.~G.~Djorgovski\altaffilmark{1},
D.~A.~Frail\altaffilmark{5},
A.~Gal-Yam\altaffilmark{6},
J.~Halpern\altaffilmark{7},
J.~A.~Holtzman\altaffilmark{8},
M.~Hunt\altaffilmark{1},
K.~Hurley\altaffilmark{9},
B.~Jacoby\altaffilmark{1},
R.~Kimble\altaffilmark{4},
S.~R.~Kulkarni\altaffilmark{1},
N.~Mirabal\altaffilmark{7},
G.~Morrison\altaffilmark{10},
E.~Ofek\altaffilmark{6},
O.~Pevunova\altaffilmark{10},
R.~Sari\altaffilmark{11},
B.~P.~Schmidt\altaffilmark{2},
D.~Turnshek\altaffilmark{3}
\& S.~Yost\altaffilmark{1}.
}

\begin{abstract}
We present multi-color light-curves of the optical afterglow of
GRB~000926.  Beginning $\sim 1.5$ days after the burst, the light-curves of
this GRB steepen measurably.  The existence of such achromatic breaks are
usually taken to be an important observational signature that the ejecta
are not expanding isotropically, but rather have a collimated jet-like
geometry.  If we interpret the data in this context, we derive an opening
angle of $5^\circ$, which reduces the energy release compared to an isotropic
model by a factor of $\sim 275$, to $1.7 \times 10^{51}$~erg.  To fit the
data with a simple jet model requires extinction along the line of sight.
The derived  $A_V$ is in the range 0.11 -- 0.82 mag, depending on the adopted
extinction  law and whether the electrons giving rise to the optical emission
are undergoing synchrotron cooling or not.  Since this is in excess of the
expected extinction from our Galaxy, we attribute this to the GRB host. We
note that this extinction is typical of a galactic disk, and therefore the
event likely took place in the disk of its host.
\end{abstract}

\keywords{cosmology: observations; gamma-rays: bursts; galaxies: ISM}


\altaffiltext{1}{Palomar Observatory, 105-24, California Institute of
Technology, Pasadena, CA, 91125.}
\altaffiltext{2}{Research School of Astronomy \& Astrophysics, Mount
Stromlo Observatory, Cotter Road, Weston, ACT, 2611, Australia.}
\altaffiltext{3}{Department of Physics and Astronomy, University of Pittsburgh, Pittsburgh, PA, 15260.}
\altaffiltext{4}{Laboratory for Astronomy and Solar Physics, Code 681, Goddard
Space Flight Centre, Greenbelt, MD, 20771.}
\altaffiltext{5}{National Radio Astronomy Observatory, P.O. Box O, Socorro, NM, 87801.}
\altaffiltext{6}{School of Physics \& Astronomy and Wise Observatory,
Tel-Aviv University, Tel-Aviv 69978, Israel.}
\altaffiltext{7}{Astronomy Department, Columbia University, 550 West 120th
Street, New York, NY, 10027.}
\altaffiltext{8}{Department of Astronomy, New Mexico State University, Box
30001, Department 4500, Las Cruces, NM, 88003-8001.}
\altaffiltext{9}{University of California Space Sciences Laboratory, Berkeley, CA, 94720.}
\altaffiltext{10}{Infrared Processing and Analysis Center 100-22, California
Institute of Technology, Pasadena, CA, 91125.}
\altaffiltext{11}{Theoretical Astrophysics 130-33, California Institute of
Technology, Pasadena, CA, 91125.}

\section{Introduction}

Multi-color light-curves of the afterglows of gamma-ray burst (GRB)
sources contain information about the evolution of the relativistic blast wave,
which results from the progenitor explosion, as it expands into the
surrounding medium.  Interpreted in the context of a theoretical afterglow
model (\cite{spn98}), the broad-band light-curve, if observed starting immediately after
the GRB through the time when the shock becomes non-relativistic, can, in
principle, provide key physical parameters, including the total energy in
the expanding ejecta, the density structure of the medium (\cite{cl99}) and
whether the ejecta are spherically-symmetric or restricted to a jet
(\cite{rhoads97}).  The optical window of the afterglow spectrum is particularly
useful for determining if the ejecta are highly collimated, since it is
generally well-sampled on timescales of hours to days when temporal decay
slope breaks due to this geometric effect become manifest.  In addition, the
effects of dust as seen through extinction are most easily observed in
multi-color optical data.

Only about a half-dozen GRB afterglows have been well-sampled in the
optical, with data of sufficient quality to test theoretical models and
provide significant constraints on the physical parameters.  In several
cases, e.g. GRB~990510 (\cite{stanek+99}, \cite{harrison+99}), GRB~991216
(\cite{halpern+00a}) and GRB~000301c (\cite{berger+00}), the optical light-curves
exhibit achromatic breaks, most easily understood as resulting from jet-like
ejecta collimated to angles of $5^{\circ}$, $6^{\circ}$ and $12^{\circ}$.
The implied degree of collimation reduces the inferred energy
release for these events by factors of 50-300.  However other events such
as GRB~970508 show no evidence for collimation in the optical (but may in
the radio --- see \cite{fwk00}), indicating that the collimation angles are
significantly larger.

In this letter, we present $BVRI$ optical monitoring of the afterglow of GRB~000926
performed by the Palomar 60-inch and 200-inch, the MDM 2.4-m telescope, and
the Wise 1.0-m telescopes, and derive the optical transient light-curve from 1
-- 7 days after the GRB.  In addition, we have obtained high-resolution HST/WFPC2
images in several bands, which allow us to properly subtract the
contribution from nearby diffuse emission, possibly associated with the GRB
host.  We have fit the multi-color data with an afterglow model and find
that observed steepening of the light-curve requires the ejecta to be
collimated in a cone.  In the context of this model, consistency of the
multi-color data and temporal decay also implies significant extinction,
likely associated with the GRB host.

\section{Observations and Data Reduction}

GRB~000926 as observed by the {\it Inter-Planetary Network} (Ulysses,
Konus-Wind, and NEAR) on 2000 Sep 26.993 UT had a duration of 25
seconds, placing it in the class of long-duration GRBs. The
position was triangulated to a relatively small error box of
approximately 35~arcmin$^{2}$ and distributed to the GRB community
0.84 days after the burst (\cite{hurley+00}). The bright $(R\sim
19.5)$ afterglow of GRB~000926 was identified by Gorosabel et~al.
(2000)\nocite{gorosabel+00} and Dall et~al. (2000)\nocite{dall+00}
from observations taken less than a day after the burst.  Spectra of
the afterglow from the Nordic Optical Telescope yielded an absorption
redshift of 2.066 (\cite{fynbo+00a}), later refined to 2.0369 $\pm $
0.0007 from Keck spectroscopy (\cite{castro+00}).

Our observations commenced with data taken by the MDM~2.4-m on Sep 28.177,
1.18 days after the burst.  A complete log of our observations and resulting
photometry can be found in Table~\ref{tab-data}.  We used the $BVRI$ filter
system for all observations save those with the Palomar 200-inch for which
observations were obtained in the ${\cal R}$ (\cite{sh93}) and Sloan $g^{\prime}$
(\cite{fukugita+96}) filters.

We calibrated twelve secondary standards in the field from observations of
Landolt (1992)\nocite{landolt92} standard star fields (Landolt fields 96, 112, 113
and 114).  These fields were observed with the Palomar 60-inch telescope on
two photometric nights.  These observations are sufficient to fit extinction terms, but not to extract a color term which we took to be zero.
From these standards and combined images of the field we also
calibrated two fainter tertiary standards for use with larger telescopes.  
For the reference star of Halpern (2000b)\nocite{halpern+00b}, we find $B = 18.468$,
$V = 17.595$, $R = 17.048$ and $I = 16.512$.  We
estimate that these calibrations are accurate to approximately 3\%.  

The Palomar 200-inch telescope photometry were transformed to the $BVRI$
system using the published transformations (\cite{sh93}, \cite{fukugita+96})
and OT colors of $(B-V)=0.61\pm 0.10$ and $(R-I)=0.75\pm 0.10$.  A 3\%
systematic error in the transformation was added in quadrature to the
statistical error in these measurements.  We compared magnitudes of field
stars measured with the Palomar 60-inch telescope with transformed magnitudes
from the 200-inch telescope photometry.  This comparison suggests that our
derived $V$ and $R$ magnitudes for the OT are accurate.

The $I$-band images display significant fringing, so the quoted formal errors
do not represent the true measurement error.  We estimated a
systematic $I$-band error of 0.09~mag by fitting a straight line through the
first four $I$-band measurements and adjusting the systematic error until
$\chi ^{2}/\mbox{DOF}=1$.

\begin{figure}[tbp]
\plotone{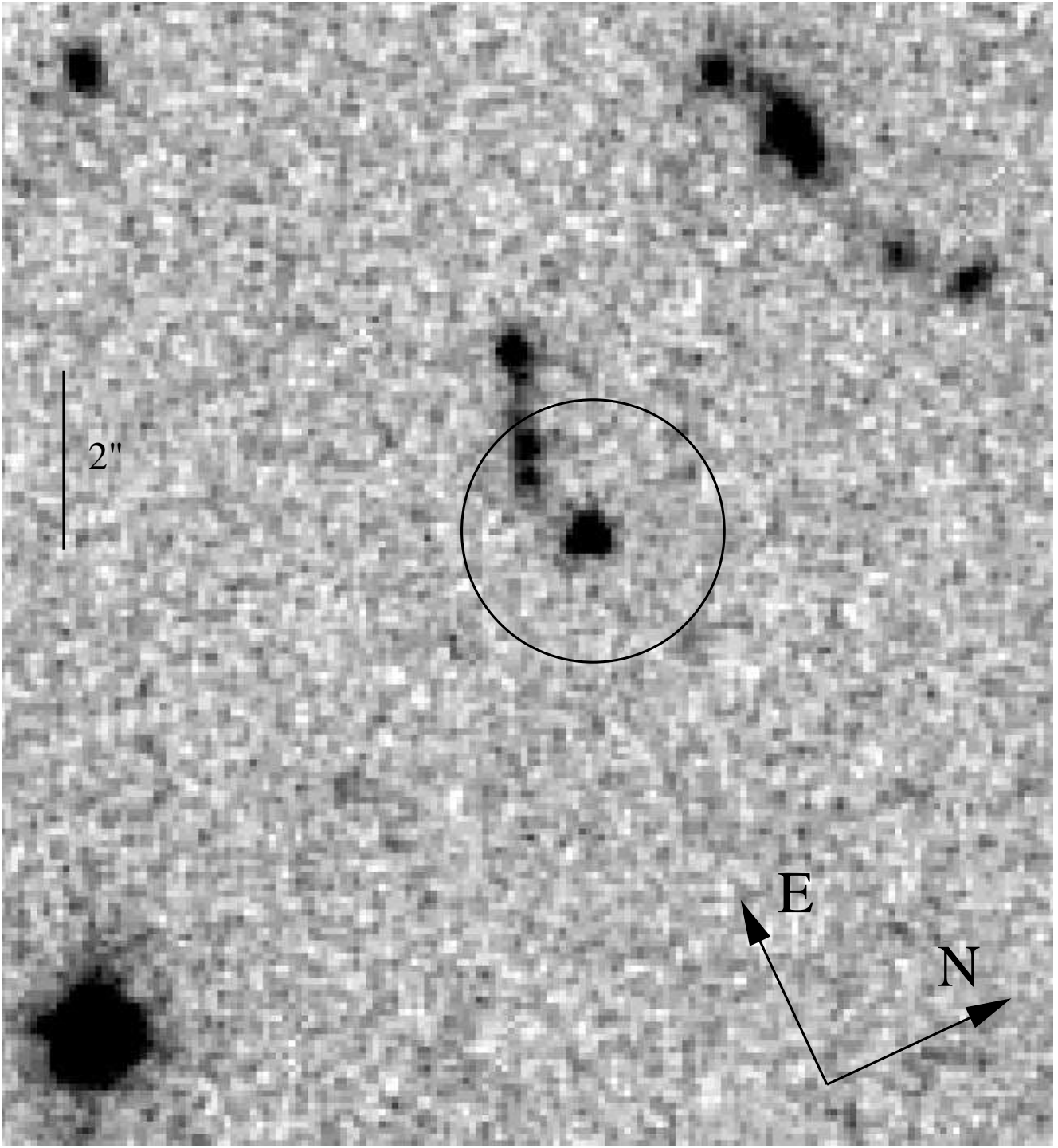}
\caption{Combined HST/WFPC2 F606W image of the GRB~000926 optical afterglow.
The extended emission approximately 1.5 arcseconds from the OT is the galaxy
contaminating the ground-based measurements.  The circle shows the aperture
(1.5 arcseconds) used for all our photometry.}
\label{fig-hst}
\end{figure}

In addition to the ground-based photometric observations, we obtained
high-resolution HST/WFPC2 images in F450W, F606W and F814W at three epochs
as part of a long-term monitoring program with HST.  The 6600 seconds (3
orbits) F450W images were combined using the STSDAS task \texttt{crrej}, while the
13200 seconds (6 orbits) F606W and F814W images were combined and cosmic-ray 
rejected using the \texttt{drizzle} technique (\cite{fh97}).  Figure~\ref{fig-hst}
displays the resultant F606W image.

\section{The Light-Curve}

Both ground-based (\cite{fynbo+00c}) and HST (Figure~\ref{fig-hst})
imaging have revealed the presence of a galaxy near the OT, which contaminates
photometry of the OT by ground-based telescopes.  Proper treatment of this
contamination is essential, since it can greatly influence the derived
late-time slope, and consequently the important physical parameters.
For example, Rol, Vreeswijk \& Tanvir (2000)\nocite{rvt00} have fit a
late-time temporal slope of $\alpha_{2} = 3.2 \pm 0.4$ for this afterglow, 
which is considerably steeper than that of other afterglows observed to date.
We therefore use our HST images to measure the contaminating galaxy flux
in a 1.5 arcsecond aperture from the OT and convert these to $BVRI$ using
Holtzman et~al. (1995)\nocite{holtzman+95}.  The results are shown in
Table~\ref{tab-gal}.  Our $R$-band measurement of the galaxy contribution is
fainter than the fit value of Rol, Vreeswijk \& Tanvir (2000)\nocite{rvt00}
of $24.2 \pm 0.3$, which may explain their steeper late-time slope.

In deriving flux values for all our ground-based data, we use a 1.5-arcsecond
aperture.  This allows us to accurately subtract the galaxy flux in a
straightforward way, using the values tabulated above.  We note that there may be an
additional compact component of the host emission not resolved by HST (which
may be observed in subsequent, scheduled observations). However, since the
light-curve shows no significant flattening, this is not likely
to be an important contribution over the interval of our observations.

\begin{figure}[tbp]
\plotone{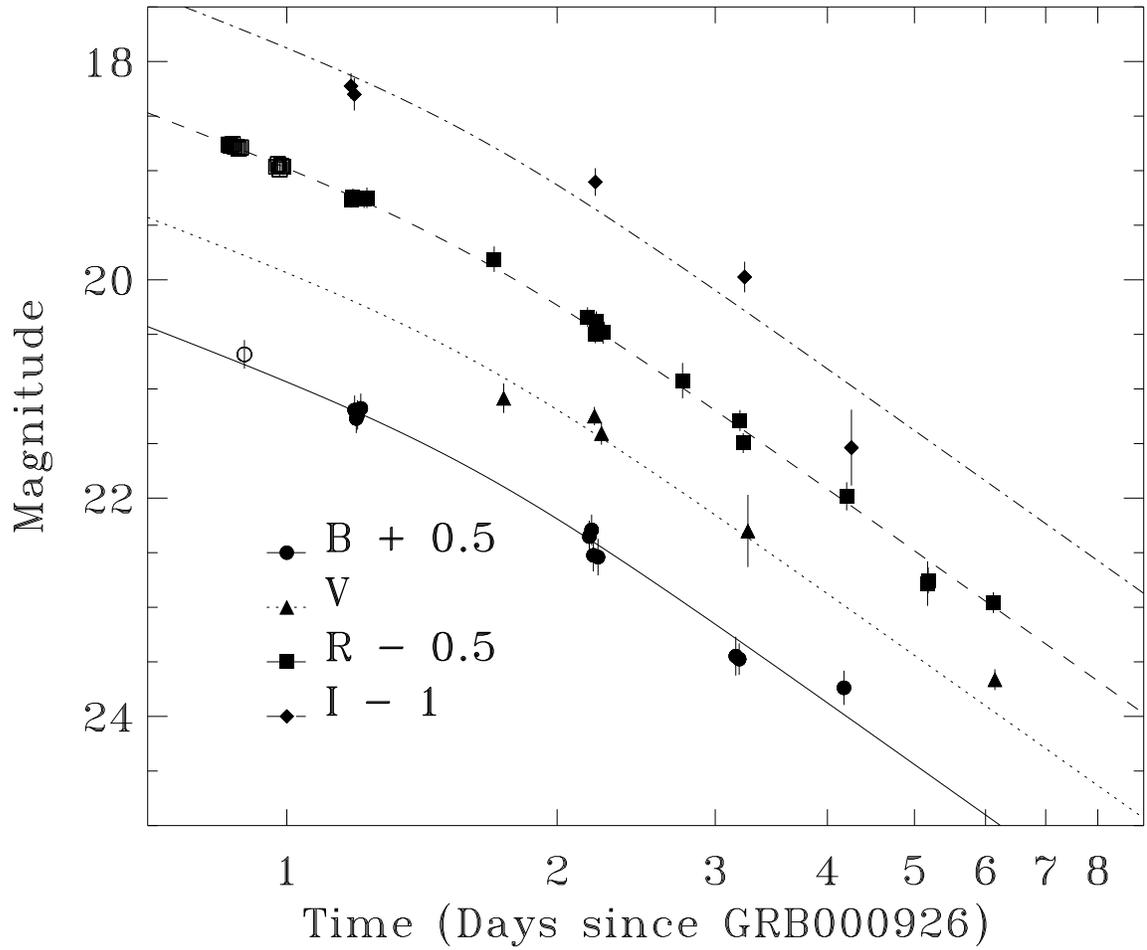}
\caption{The BVRI light-curve of GRB~000926.  Filled points are data
presented in this work; hollow points ($t < 1$ day) are from Hjorth
et~al. (2000) and Fynbo et~al. (2000b).  The measurements have had the
contaminating galaxy flux subtracted.  The solid line shows the best fit
to Equation~\ref{eq-function}.}
\label{fig-plot}
\end{figure}

Since the aperture size used in measurements reported through the GRB
Coordinate Network Circulars (GCN)\footnote{GCN circulars can be accessed from
http://gcn.gsfc.nasa.gov/gcn/gcn3\_archive.html} is generally unspecified
and variable, the amount of contamination by the galaxy in each measurement
cannot be determined.  Consequently, we include only measurements taken within
one day of the GRB (\cite{hjorth+00}, \cite{fynbo+00b}), in addition to the
measurements presented in this paper in constructing the light-curve.  These
data are important for constraining the early-time temporal slope, and at
these times, the OT is bright and the contamination by the galaxy is
negligible.  The measurements from the GCN were re-calibrated using our
secondary standards.  We correct all measurements for foreground
Galactic extinction using $E_{B-V}=0.023$ mag from Schlegel, Finkbeiner \&
Davis (1998)\nocite{sfd98}. In Figure~\ref{fig-plot} we display the OT light-curve
in which the contribution from the host galaxy has been subtracted and the Galactic reddening
has been accounted for.
A single power-law temporal decay is clearly excluded, with a probability that
it fits the data of $3\times 10^{-6}$.  

In order to characterize the
light-curve, we have fit it to the functional form
(\cite{beuermann+99})
\begin{equation}
F(t,\nu )=F_{0}\ \nu ^{\beta }\ \left[
 (t/t_{*})^{-\alpha_{1}s}+(t/t_{*})^{-\alpha_{2}s}\right] ^{-1/s}. 
\label{eq-function}
\end{equation}
This function
has no physical significance, but provides a simple and general parametric
description of the data, allowing for a gradual break in the afterglow decay. In this
function, $\alpha_{1}$ and $\alpha_{2}$ are the early and late time
asymptotic temporal slopes respectively, $t_{*}$ is the time of the
temporal slope break, $\beta $ is the spectral slope, and $s$ is a parameter
that determines the sharpness of the transition.  We leave the break sharpness
as a free parameter, since there is disagreement over its theoretical value
(eg., \cite{kp00}).

We first fit equation \ref{eq-function} without any constraints, applying 5\%
systematic error (in addition to the errors given in Table~\ref{tab-data}
above) to all measurements to reflect uncertainties in zero point calibrations
for the different telescopes and in the conversion of WFPC2 magnitudes to $BVRI$.
This form fits well, with $\chi^{2}/\mbox{DOF} = 48.5/45 $ and the fit
parameters $t_{*}=1.79\pm 0.15$ days, $\alpha_{1}=-1.48\pm 0.10$,
$\alpha_{2}=-2.302\pm 0.082$, and $\beta =-1.522\pm 0.066$, where the
errors do not reflect covariance between the parameters.  The best fit
value for the break sharpness value is $s = 15$, but is not well constrained,
due to the lack of early-time data.  Figure~\ref{fig-plot} shows this fit
overplotted on the data points. 

\section{Interpretation}

We have demonstrated that the observed break in the light-curve is consistent
with being achromatic, since the parameter $t_{*}$ is independent of frequency.
This frequency-independent steepening of the optical light-curve is most
easily interpreted as due to collimated, or jet-like ejecta.  Once the
Lorentz factor of the ejecta falls below the inverse of the opening
angle of the jet the light-curve steepens due to geometric effects, as well
as due to the sideways expansion of the ejecta (Rhoads 1997, 1999\nocite{rhoads97}
\nocite{rhoads99}).  Interpreted in this context, the early and late-time light-curve
slopes, the optical spectral index, and the time of the transition constrain
the index of the electron spectral energy distribution, $p$, the jet opening angle and the total
energy of the afterglow.

We now adopt the simple model developed by Sari, Piran and Halpern
(1999)\nocite{sph99}.  This model predicts the temporal and spectral evolution
of synchrotron radiation from a jet expanding relativistically in a constant
density medium. The early- and late-time temporal slopes and the optical
spectral slope are determined by the electron spectral index, $p$, and the
break time is determined by the jet opening angle.  Optical data alone do not
have sufficient frequency coverage to locate all of the afterglow spectral
breaks.  Specifically, with the optical light-curve we cannot constrain the
position of the cooling break, $\nu_{c}$, and we must consider two cases: 1)
$\nu_{c}$ is blue-ward of the optical (henceforth referred to as ``case B''); and
2) $\nu_{c}$ is red-ward of the optical (henceforth ``case R'').

We find that when we fit light-curves from all optical bands simultaneously (linking
the spectral slope and the two temporal decay
slopes) using the theoretical predictions of the model, we cannot produce an
acceptable fit to the data.  Our $\chi^{2}$ values
of 195 and 83 for 47 degrees of freedom for case B and case R respectively
correspond to a probability that the model describes
the data of less than $2 \times 10^{-4}$.  Clearly, the observed optical
spectral index is inconsistent with the model, being too steep for the value
of $p$ determined from the temporal decay slopes.

This problem can be resolved if we include the effect of extinction in the
host galaxy of the GRB, which can modify the spectral index.  This explanation
is consistent with the strong equivalent widths of absorption lines observed
in spectra of this afterglow from the Keck telescope (\cite{castro+00}). The
appropriate extinction law is, however, unknown and unconstrained by our data,
so to determine the source-frame $A_{V}$ we consider several possibilities. We
allow for extinction laws corresponding to young star-forming regions (such as
the Orion Nebula), the Milky Way, the LMC and the SMC by using the Cardelli,
Clayton \& Mathis (1989)\nocite{ccm89} and the Fitzpatrick \& Massa
(1988)\nocite{fm88} extinction curves, with the smooth joining calculated by
Reichart (1999)\nocite{r99}.

Including extinction provides an acceptable fit to the multi-band data
for both cases, with an electron energy spectral index $p=2.38 \pm 0.15$
which is consistent with that found for other afterglows.  For case B,
the derived $A_{V}$ values range from 0.82 mag for the
Milky Way extinction law to 0.28 mag and 0.25 mag for the LMC and the
SMC extinction laws ($\chi^{2}\approx\ 50$ for 46 degrees of freedom), with
a break time of $t_{*} = 1.45\pm 0.14$ days.
For case R, $A_{V}$ is 0.36/0.12/0.11 for Milky Way/LMC/SMC
extinction, with a break time of $t_{*} = 1.60\pm 0.13$ days.  In both cases,
an extinction law corresponding to a young star-forming region does not fit
the data, since it is `grey' in the source-frame UV.  The parameter $p$ is
insensitive to the extinction law and the position of the cooling
break to within the quoted error.  We calculate the corresponding jet
half-opening angle using Sari, Piran \& Halpern (1999)\nocite{sph99}
to be $\theta_{0}\sim 5^{\circ}\ n_{1}^{1/8}$, where $n_{1}$ is the
density of the ISM, in units of cm$^{-3}$.

\section{Conclusions}

Our well-sampled multi-color light-curve of the afterglow of GRB~000926 is
well-described by a physical model where the ejecta are collimated in a jet.
The degree of collimation reduces the inferred isotropic radiated energy
of the GRB (\cite{bloom+01}) by a factor of 275, to $1.7\times
10^{51}\ n_{1}^{1/4}$ ergs. This inferred energy release is typical of events
observed to-date.  Furthermore, we find that in order to properly fit the
light-curve of this afterglow extinction is required. Assuming the extinction
is at the measured redshift of $z = 2.0369$ (\cite{castro+00}), we can exclude
an extinction law corresponding to a young star-forming region, and we find an
$A_{V}$ ranging from 0.11 -- 0.82 mag, depending on the assumed curve and
on the cooling regime.  This value exceeds the expected extinction from our
own Galaxy, and is likely due to the host galaxy of the GRB.

\acknowledgements

We thank the staff of the Palomar, Keck, MDM and Wise Observatories,
and also L. Cowie, A. Barger, R. Ellis, C. Steidel and B. Madore for
their assistance in obtaining observations.  We thank E. Mazets and
the Konus team for the IPN data.  FAH acknowledges support from a
Presidential Early Career award.  SRK, SGD and JPH thank NSF for support of
their ground-based GRB programs.  KH is grateful for Ulysses support
under JPL contract 958056 and NEAR support under NAG5-9503.

\clearpage

\begin{deluxetable}{cccc}
\footnotesize
\tablecolumns{4}
\tablewidth{0pc}
\tablecaption{Ground-based measurements of the GRB~000926 optical afterglow
made as a part of this work.\label{tab-data}}
\tablehead{\colhead{Date (2000, UT)} & \colhead{Filter} & \colhead{Magnitude} & \colhead{Telescope}}
\startdata
Sep 28.183 &  $	B  $  &	 $20.890 \pm\ 0.038  $ &	MDM 2.4-m	\nl
Sep 28.188 &  $	B  $  &	 $20.967 \pm\ 0.039  $ &	MDM 2.4-m	\nl
Sep 28.192 &  $	B  $  &	 $20.934 \pm\ 0.043  $ &	Palomar 60-inch	\nl
Sep 28.202 &  $	B  $  &	 $20.874 \pm\ 0.044  $ &	Palomar 60-inch	\nl
Sep 29.165 &  $	B  $  &	 $22.039 \pm\ 0.071  $ &	Palomar 60-inch	\nl
Sep 29.178 &  $	B  $  &	 $21.979 \pm\ 0.057  $ &	MDM 2.4-m	\nl
Sep 29.188 &  $	B  $  &	 $22.208 \pm\ 0.074  $ &	MDM 2.4-m	\nl
Sep 29.214 &  $	B  $  &	 $22.22 \pm\ 0.11    $	  &	Palomar 60-inch	\nl
Sep 30.155 &  $	B  $  &	 $23.10 \pm\ 0.12    $	  &	Palomar 60-inch	\nl
Sep 30.183 &  $	B  $  &	 $23.126 \pm\ 0.067  $ &	MDM 2.4-m	\nl
Oct 1.166  &  $	B  $  &	 $23.373 \pm\ 0.091  $ &	Palomar 60-inch	\nl
			  		      
Sep 28.737 &  $	V  $  &	 $21.25 \pm\ 0.12    $	  &	Wise 1.0-m	\nl
Sep 29.194 &  $	V  $  &	 $21.416 \pm\ 0.063  $ &	Palomar 60-inch	\nl
Sep 29.234 &  $	V  $  &	 $21.573 \pm\ 0.087  $ &	Palomar 60-inch	\nl
Sep 30.255 &  $	V  $  &	 $22.45 \pm\ 0.33    $	  &	Palomar 60-inch	\nl
Oct 3.138  &  $	V\tablenotemark{a}  $  &    $23.726 \pm\ 0.077$  &	Palomar 200-inch\nl
			  		      
Sep 28.173 &  $	R  $  &	 $19.918 \pm\ 0.020  $ &	MDM 2.4-m	\nl
Sep 28.178 &  $	R  $  &	 $19.890 \pm\ 0.019  $ &	MDM 2.4-m	\nl
Sep 28.212 &  $	R  $  &	 $19.917 \pm\ 0.033  $ &	Palomar 60-inch	\nl
Sep 28.221 &  $	R  $  &	 $19.902 \pm\ 0.062  $ &	Palomar 60-inch	\nl
Sep 28.695 &  $	R  $  &	 $20.461 \pm\ 0.093  $ &	Wise 1.0-m	\nl
Sep 29.155 &  $	R  $  &	 $20.985 \pm\ 0.054  $ &	Palomar 60-inch	\nl
Sep 29.198 &  $	R  $  &	 $21.139 \pm\ 0.043  $ &	MDM 2.4-m	\nl
Sep 29.204 &  $	R  $  &	 $21.028 \pm\ 0.064  $ &	Palomar 60-inch	\nl
Sep 29.207 &  $	R  $  &	 $21.094 \pm\ 0.041  $ &	MDM 2.4-m	\nl
Sep 29.243 &  $	R  $  &	 $21.118 \pm\ 0.083  $ &	Palomar 60-inch	\nl
Sep 29.752 &  $	R  $  &	 $21.55 \pm\ 0.15    $	  &	Wise 1.0-m	\nl
Sep 30.189 &  $	R  $  &	 $21.906 \pm\ 0.065  $ &	Palomar 60-inch	\nl
Sep 30.216 &  $	R  $  &	 $22.103 \pm\ 0.057  $ &	MDM 2.4-m	\nl
Oct 1.195  &  $	R  $  &	 $22.56 \pm\ 0.11    $	  &	Palomar 60-inch	\nl
Oct 2.164  &  $	R  $  &	 $23.26 \pm\ 0.19    $	  &	Palomar 60-inch	\nl
Oct 2.172  &  $	R  $  &	 $23.235 \pm\ 0.095  $ &	MDM 2.4-m	\nl
Oct 3.113  &  $	R\tablenotemark{b}  $  &    $23.402 \pm\ 0.063$  &	Palomar 200-inch\nl
			  		      
Sep 28.172 &  $	I  $  &	 $19.359 \pm\ 0.036  $ &	Palomar 60-inch	\nl
Sep 28.182 &  $	I  $  &	 $19.435 \pm\ 0.096  $ &	Palomar 60-inch	\nl
Sep 29.199 &  $	I  $  &	 $20.230 \pm\ 0.057  $ &	Palomar 60-inch	\nl
Sep 30.228 &  $	I  $  &	 $21.079 \pm\ 0.083  $ &	Palomar 60-inch	\nl
Oct 1.247  &  $	I  $  &	 $22.51 \pm\ 0.33    $	  &	Palomar 60-inch	\nl
\enddata
\tablenotetext{a}{This observation was made using a Sloan $g'$ filter (\cite{fukugita+96}).}
\tablenotetext{b}{This observation was made using a $\Re$ filter (\cite{sh93}).}
\end{deluxetable}

\clearpage

\begin{deluxetable}{cc}
\footnotesize
\tablecolumns{2}
\tablewidth{0pc}
\tablecaption{HST/WFPC2 measurements of the contaminating galaxy flux within
a 1.5 arcsecond aperture centred on the OT.\label{tab-gal}}
\tablehead{\colhead{Band} & \colhead{Magnitude}}
\startdata
$ B $ & $ 26.23 \pm\ 0.50 $ \nl
$ V $ & $ 26.09 \pm\ 0.16 $ \nl
$ R $ & $ 25.19 \pm\ 0.17 $ \nl
$ I $ & $ 24.50 \pm\ 0.11 $ \nl
\enddata
\end{deluxetable}

\clearpage

\clearpage

\clearpage

\end{document}